\documentclass[%
 reprint,
superscriptaddress,
 amsmath,amssymb,
 aps,
]{revtex4-2}

\usepackage{graphicx}
\usepackage{float}
\usepackage{textcomp}
\usepackage{gensymb}
\usepackage{xcolor}

\usepackage{hyperref}
\hypersetup{colorlinks=true,citecolor=violet, linkcolor=orange, filecolor=magenta, urlcolor=cyan,}

\begin{document}

\title{Spatial, Spectral and Temporal Response of High Intensity Laser Plasma Mirrors--Direct Observation of the Ponderomotive Push}



\author{Sk Rakeeb}
\affiliation{Tata Institute of Fundamental Research, Mumbai, India, 400005}

\author{Animesh Sharma}
\affiliation{Indian Institute of Technology Delhi, India, 110016}

\author{Sagar Dam}
\affiliation{Tata Institute of Fundamental Research, Mumbai, India, 400005}

\author{Ameya Parab}
\affiliation{Tata Institute of Fundamental Research, Mumbai, India, 400005}

\author{Amit D. Lad}
\email{Present address: Center for Quantum Science and Technologies, Indian Institute of Technology Mandi, Kamand, Mandi, India, 175075}
\affiliation{Tata Institute of Fundamental Research, Mumbai, India, 400005}

\author{Yash.M. Ved}
\affiliation{Tata Institute of Fundamental Research, Mumbai, India, 400005}

\author{Amita Das}
\affiliation{Indian Institute of Technology Delhi, India, 110016}

\author{G. Ravindra Kumar}
\email{Corresponding author: grk@tifr.res.in} 
\affiliation{Tata Institute of Fundamental Research, Mumbai, India, 400005}


\date{\today}

\begin{abstract}
Plasma-based optics have emerged as a powerful platform for manipulating and amplifying ultra-intense laser pulses. However, the inherently nonlinear and dynamic nature of plasma leads to significant spatial, spectral, and temporal modulations when driven at relativistic intensities. These modifications can dramatically alter the structure of the reflected laser pulses, posing challenges for their use in applications such as vacuum ultraviolet (VUV) and X-ray generation, as well as relativistic particle acceleration. Comprehensive, multidimensional diagnostics are essential to accurately characterize these so-called `plasma mirrors' (PMs).
We present a direct, \textit{in situ}, and first-of-its-kind measurement of the three-dimensional evolution of a plasma surface during femtosecond laser irradiation, achieved through the simultaneous analysis of the wavefront, spectral, and temporal characteristics of the reflected light. The measurements reveal real-time three-dimensional surface deformations driven by the well-known ponderomotive force arising from spatial and temporal intensity gradients, in good agreement with three-dimensional particle-in-cell (3D-PIC) simulations. In addition, the plasma mirror induces substantial modifications to the pulse spectrum and temporal profile, leading to the emergence of spatio-temporal couplings, more pronounced in higher-order harmonics.

\end{abstract}


\maketitle

\section{Introduction}
 Recent advancements in laser technology facilitate the study of the interaction between high-power, ultrahigh contrast femtosecond pulses with solid-density matter \cite{drake2006introduction, kaw2017nonlinear, ravindra2009intense,rocca2024ultra}. This opens up avenues in high-energy density science, with potential applications in high-brightness vacuum ultraviolet (VUV) and X-ray sources \cite{constant1999optimizing,dromey2007bright}, attosecond optics \cite{thaury2007plasma}, tabletop ultrafast X-ray imaging \cite{kneip2010bright,kneip2008observation} and ultrafast diffraction \cite{gaffney2007imaging}. These developments are critically dependent on the temporal and spatial characteristics of femtosecond pulses \cite{jolly2020spatio}. Unlike picosecond and nanosecond pulses, where interaction occurs in a plasma that expands below solid density, ultrahigh contrast femtosecond pulses can induce near-instantaneous excitation of solid-density matter \cite{murnane1991ultrafast}. In addition short-pulse, high-intensity lasers can create conditions where quantum electrodynamics (QED) effects \cite{marklund2009quantum} and relativistic effects \cite{umstadter2001review} become significant. In a seminal paper, Wilks et al.~\cite{wilks1992absorption} used PIC simulations to show that the extreme pressure (of the order of several hundred megabars) exerted on the plasma surface leads to surface deformation, resulting in various instabilities~\cite{tatarakis2003propagation}. Once the plasma is formed, the laser-plasma interaction is predominantly governed by the critical density surface, which acts as a reflective and absorptive interface—referred to as the `plasma mirror' (PM).
 The behavior of this PM, especially under relativistic conditions, strongly modulates the incident laser pulse. A moving relativistic plasma mirror can shift the frequency, reshape the temporal profile, and alter the spatial structure of the reflected light~\cite{thaury2007plasma}.

Accurate characterization of the PM \textit{i.e.} its spatial deformation, spectral response, and temporal evolution is therefore crucial for understanding the laser-plasma coupling mechanism. At relativistic intensities, the ponderomotive force (PMF) becomes the primary driver of plasma dynamics. This nonlinear force, arising from the spatial and temporal inhomogeneities of the laser intensity, pushes electrons away from regions of high intensity. The PMF governs a broad spectrum of physical processes, including nonlinear self-action of the laser pulse, relativistic surface denting, and the generation of fast electrons and ions~\cite{kodama2004plasma, macchi2005laser}.

Phenomena such as radiation pressure acceleration (RPA)~\cite{esirkepov2004highly,robinson2008radiation,macchi2013ion,schlegel2009relativistic}, high-harmonic generation (HHG)~\cite{thaury2007plasma}, and laser–plasma absorption mechanisms including resonance absorption~\cite{gibbon1992collisionless} are strongly influenced by the curvature and motion introduced at the plasma surface by the ponderomotive force. In these regimes, even subtle variations in the shape of the plasma mirror can significantly alter the local incidence angle, field enhancement, and energy-coupling pathways, thereby affecting the efficiency and spectral characteristics of the resulting ion/electron beams, harmonics, or absorbed energy.

Over the past decades, considerable progress has been made in diagnosing these ultrafast plasma dynamics. Measurement techniques based on spectral signatures-such as Doppler shifts, and spectral modulations have provided valuable access to plasma expansion velocities and energy deposition mechanisms~\cite{liu1992competition,malka1996experimental,adak2015terahertz,dulat2022subpicosecond,mondal2010doppler}. Likewise, HHG-based diagnostics have enabled high-resolution probing of plasma surfaces through the angular emission patterns and spectral content of generated harmonics~\cite{vincenti2014optical,chopineau2021spatio}.

\begin{figure*}[htbp!]
        \centering
        \includegraphics[width=1\linewidth]{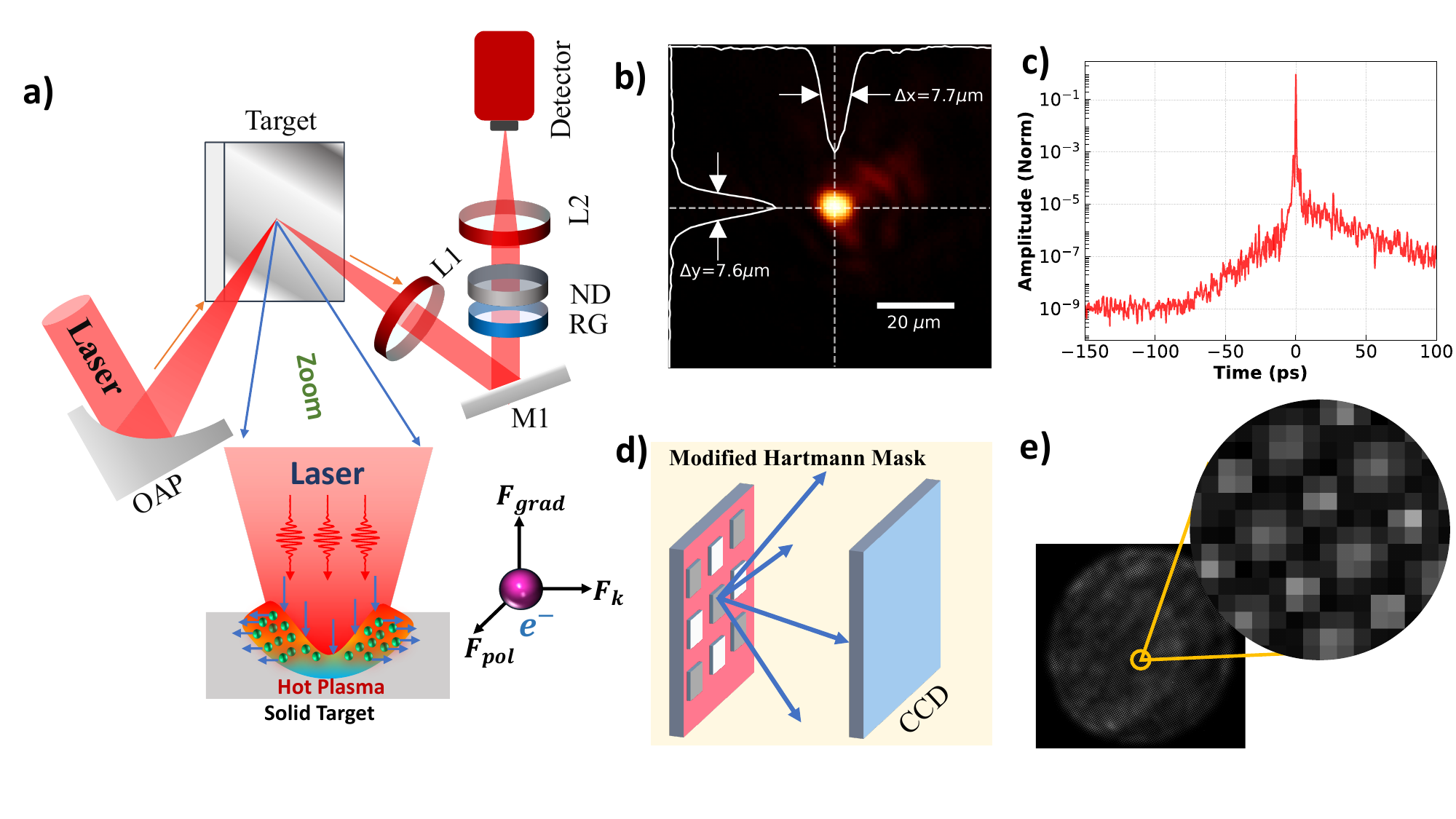}
        
        \caption{\textbf{a)} Schematic of the experimental setup for measuring laser effects on electrons with a QWLSI wavefront sensor as the detector. 
        The figure illustrates the ponderomotive push caused by the intense laser and shows various forces acting on an electron. The PMF on an electron has three components along three directions, as illustrated. $F_{pol}$ is directed along the polarization direction, $F_{k}$ is in the laser propagation direction, and $F_{grad}$ is along the gradient of the average intensity of the field.
        \textbf{b)} Measured focal spot in vacuum. \textbf{c)} Laser contrast measured using a third-order nonlinear process. \textbf{d)} Diagram of a QWLSI wavefront sensor, which comprises a cross grating and a CCD. \textbf{e)} Example of a raw and zoomed-out portion of an interferogram as measured by the QWLSI wavefront sensor.}

        \label{fig:Exp_Setup}
    \end{figure*}

Nevertheless, spectrum-based methods remain fundamentally one-dimensional, typically yielding only integrated velocity information without resolving the actual geometry of the curved plasma surface. HHG-based imaging can retrieve two-dimensional information but provides an indirect measure of the surface shape, relying on inversion of the harmonic emission rather than direct observation of the interface itself. Thus, neither approach offers a complete, model-independent reconstruction of the instantaneous curvature and spatial profile of the plasma mirror.

Because of these constraints, researchers frequently turn to analytical models and numerical simulations to interpret experimental observations~\cite{wilks1992absorption,macchi2005laser,quesnel1998theory,fedorov1997ponderomotive,schumacher2011shaped,kingham2001phase}. While these tools have been very useful for developing theoretical insight, they depend critically on assumed initial conditions, density gradients, surface motion, or laser parameters. In reality, the plasma surface evolves on femtosecond to picosecond timescales in a highly nonlinear manner \cite{wilks1992absorption}, making it challenging for simplified models to capture the full complexity of the interaction.

These limitations highlight the growing need for diagnostics capable of directly measuring the three-dimensional geometry and evolution of relativistic plasma mirrors. Such measurements would not only validate existing theoretical frameworks but also provide new input for advancing our understanding of laser-driven acceleration, harmonic generation, and absorption mechanisms under extreme conditions.

In this work, we report the first direct, \textit{in situ} three-dimensional surface mapping of a plasma mirror during femtosecond pulse interactions by measuring the wavefront of the reflected light from the plasma. The wavefront carries the information of surface depth and provides a direct and easy measure of the plasma surface. The deformation has been measured for different intensities and has been shown to agree with the theoretical model developed in \cite{vincenti2014optical}. In addition we measured the spectrum and temporal profile of the reflected pulse to gain deeper insight into the underlying plasma dynamics. The reflected pulse undergoes significant modifications in both the spectral and temporal domains. For comparison, an `extreme' contrast second harmonic pulse \cite{aparajit2023femtosecond} was employed to examine the spatial and spectral properties of the reflected beam. Our high-fidelity 3D3V and 2D3V particle-in-cell (PIC) simulations successfully model the laser-induced depression of plasma density surfaces caused by the PMF. 
\section{Experiment}

\begin{figure*}[htbp!]
        \centering
        \includegraphics[width=1\linewidth]{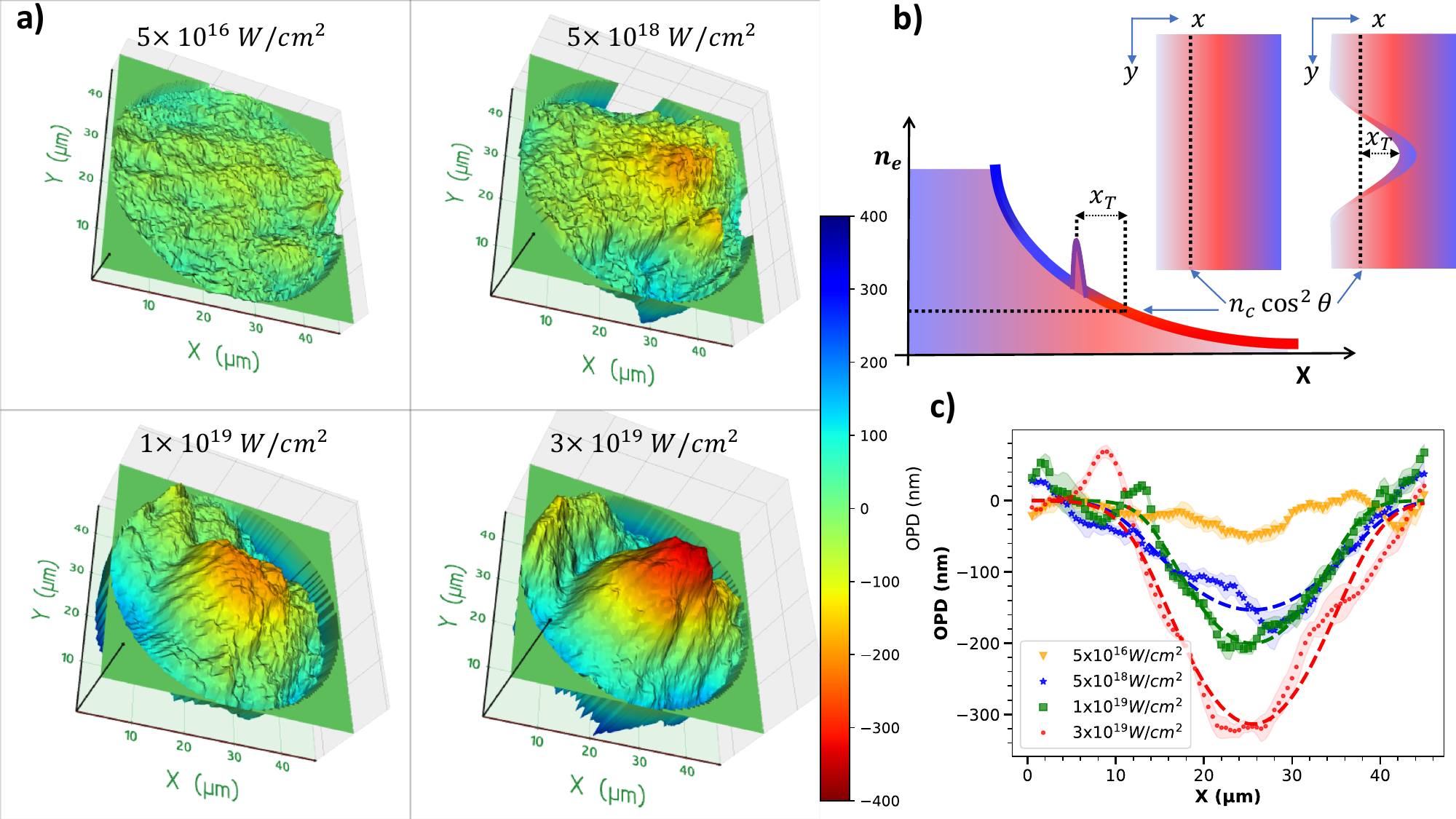}
        
        \caption{\textbf{a)} Wavefront of the reflected laser at different intensities for Setup-A. These are real time 3D surface deformations measured. X and Y represents the spatial dimension and Z is the phase plotted in terms of OPD in vacuum. (Note: for better visualization, the plot has been flipped upside down along the Z-axis) \textbf{b)} Top left sketch represents two dimensional density profile at low intensity, just start of the pulse, as the intensity increases, the plasma surface gets deformed, piling up of electrons and ions happens, represented by the top right figure. The bottom figure shows the exponential density profile in the x direction. The pulse initially reflects from $n_c cos^2(\theta)$ where $\theta$ is the incidence angle with respect to the target normal. The cartoon also shows the density bump created by the laser pulse and the total surface deformation.  \textbf{c)} Line outs taken at each intensities. Line curves represent the corresponding surface deformation obtained from the theoretical calculation. Shows the experimental data and the model is in good agreement. }
        \label{fig:Setup-A_wavefronts}
    \end{figure*}

The experiments were performed using the 150~TW laser system at the Tata Institute of Fundamental Research (TIFR), Mumbai~\cite{chatterjee2014high}. A Ti:Sapphire laser (30~fs, 800~nm) with a high picosecond contrast of $10^{-9}$ (at tens of ps) and capable of producing relativistic intensities was employed to generate overdense plasma on various solid targets. 

The laser beam was focused using an f/3 off-axis dielectric-coated parabolic mirror, enabling the achievement of different peak intensities and operation at various angles of incidence (AOI). The laser light reflected from the plasma target was collected and directed outside the vacuum chamber. The reflected beam was imaged onto a wavefront-sensing device (Phasics SID-4), as shown in Fig.~\ref{fig:Exp_Setup}(a). In addition, a fraction of the beam was sent to an Ocean Optics HR2000 spectrometer for spectral measurements and to a single-shot temporal characterization device, GRENOUILLE (Frequency-Resolved Optical Gating)~\cite{trebino2000frequency}. Throughout the experimental campaign, several experimental configurations, summarized in Table~\ref{table1}, were employed to investigate the spatial, spectral, and temporal responses of high-intensity plasma mirrors.

\begin{table}[htb]
    \centering
    \caption{\label{table1} Different experimental scenarios applied}
    \resizebox{\columnwidth}{!}{
    \begin{tabular}{|l| l| l| l|l|l|l|}
    \hline
    \hline
    Setup &   Laser \( \lambda_{L} \)   & Beam Profile & Target       &Focal Spot & Contrast & AOI\\ 
    \hline
    A    & 800 nm  & Gaussian & Plane BK7    & 7 \( \mu m\)  & $10^{-9}$  & 45\degree\\
    \hline
    B    & 800 nm  & Gaussian &  Sub-$\lambda$ Grating    & 7 \( \mu m\) & $10^{-9}$  & 23.5\degree\\
    \hline
    C    & 800 nm  & Flat-top &   Plane BK7    & 90 \( \mu m\) & $10^{-9}$  & 45\degree\\
    \hline
    D    & 400 nm  & Gaussian &   Plane BK7    & 12 \( \mu m\) & $10^{-18}$\footnote{Expected}   & 45\degree\\
    \hline
    \hline
    \end{tabular}
    }
\end{table}
We use a high definition wavefront measuring device also known as QuadriWave Lateral Shearing Interferometer (QWLSI), invented by Primot et al. in 2000\cite{primot2000extended}\cite{primot1995achromatic}. The QWLSI is composed of a two dimensional transmission grating, a phase checkerboard, and a CCD device. Four replicas of the sample wavefront to be measured are produced and the interference pattern is recorded on a CCD camera. The period of the phase checkerboard is adjusted such that 4 waves are generated from 4 first-order diffraction orders of the incoming wave (see Fig.~\ref{fig:Exp_Setup}d). The wavefront is retrieved from the interferogram by properly masking the spatial frequencies in the Fourier domain and integrating the gradients along the transverse axis.  The combination of the grating and phase checkerboard is called a modified Hartmann mask (MHM) due to its similarity with the traditional Hartmann sensors. As QWLSI wavefront devices are very sensitive (10 nm absolute RMS) and have a high spatial resolution (27 $\mu m$), these devices are routinely used in adaptive optics setups for wavefront correction in high-intensity lasers where the wavefront becomes heavily distorted by the different amplification stages \cite{yoon2021realization}. They also are used extensively in biology for phase microscopy\cite{baffou2023wavefront}.

The phase obtained from the sensor was converted into the Optical Path Difference (OPD) corresponding to free-space propagation in vacuum, using the relation $OPD = \Phi\lambda/{2\pi}$, where $\Phi$ is the phase and $\lambda$ is the wavelength. Phase measurements were conducted at different target intensities, creating distinct regimes of laser-matter interaction. At each intensity level, 10 laser shots were taken, and the phases were averaged to reduce numerical error.   

\section{Results and Discussion}
\begin{figure*}[htbp!]
        \centering
        \includegraphics[width=1\linewidth]{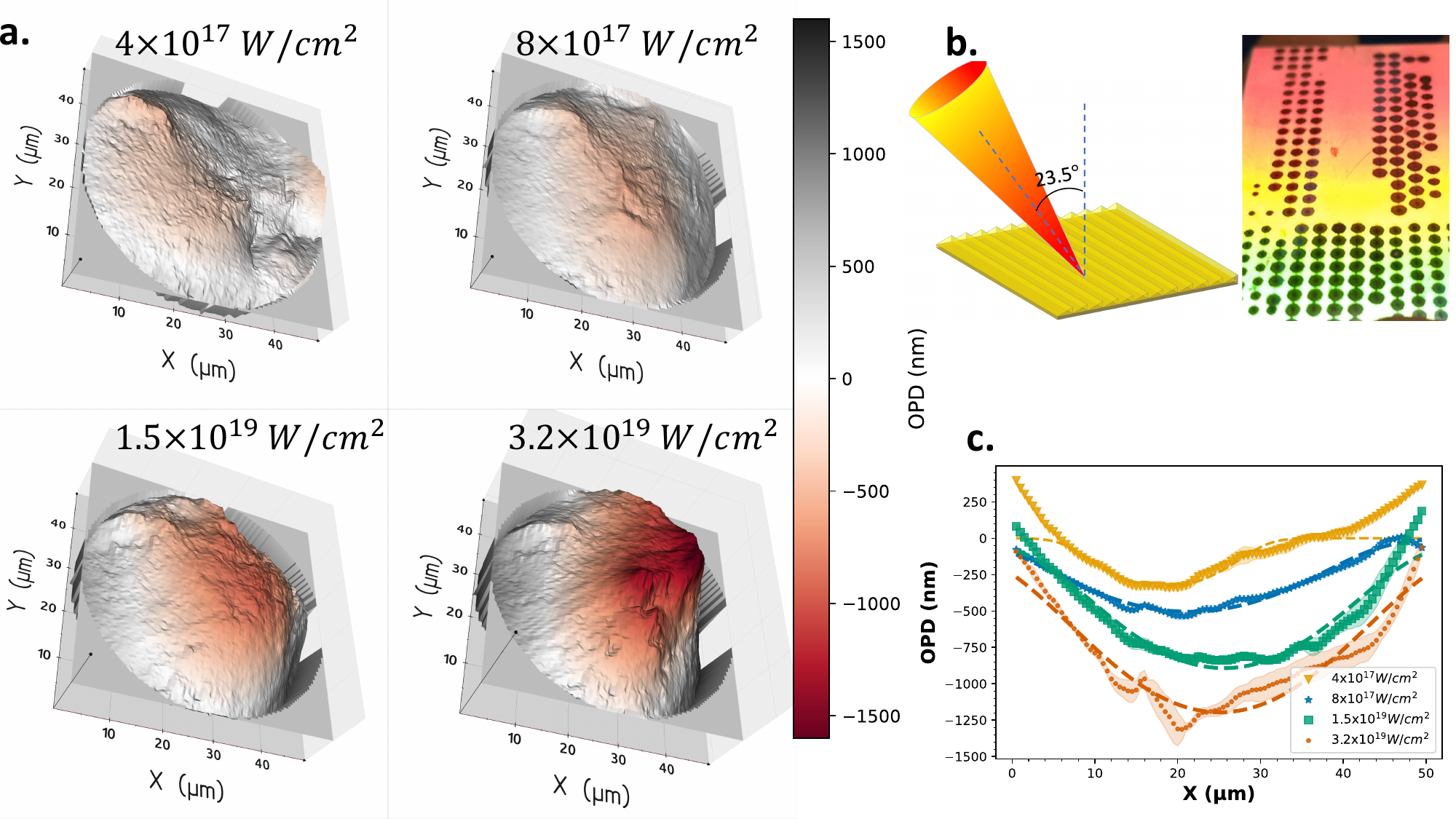}
        
        \caption{\textbf{a)} Wavefront of the reflected laser at different intensities from sub-$\lambda$ grating [Setup-B]. (Note: for better visualization, the plot has been flipped upside down along the Z-axis.) \textbf{b)} Left: Schematic of the laser incidence on the grating, Right: image of laser shots taken on the grating. \textbf{c)} The linecut of the wavefronts represented on \textbf{a}.  Line curves represent the corresponding surface deformation obtained from the theoretical calculation. }

        \label{fig:grating_wf_235}
\end{figure*}

It is well known that ponderomotive force (PMF) arises from spatial intensity gradients in an electric field  \cite{panofsky2012classical, greiner2012classical}. In the non relativistic laser case, the PMF is a time-averaged force having direction opposite to that of the intensity gradient, and is independent of the particle charge. In general, the PMF \textit{$F_P$} can be expressed as $F_P = - \frac{\partial T_{ik}}{\partial x_k}$ where summation is applied on the repeated index and $T_{ik}$ is the Maxwell stress tensor\cite{gamaly1993ponderomotive}. In the relativistic regime, although the main features of the non-relativistic PMF hold (\textit{i.e.} electrons expelled form high intensity regions and average motion is independent of the laser polarization), the force mainly depends on the negative gradient of the effective mass\cite{bauer1995relativistic}. Relativistic treatment has been done theoretically for a focused intense laser\cite{quesnel1998theory}. Also calculations have been carried out for electrons in a weakly inhomogeneous light field \cite{bituk1999relativistic} which typically is the case for a laser focused with a focusing element having focal length much larger than the laser wavelength($f>>\lambda$). 
In this case the PMF on a electron has three components along three directions as shown in (Fig.~\ref{fig:Exp_Setup}a), $F_{pol}$ is directed along the polarization direction, $F_{k}$ is in the laser propagation direction and $F_{grad}$ is along the gradient of the average intensity of the field.  

For a laser with intensity \textit{I}, the pressure exerted is given by $P = (1+R)I/c$, where \textit{R} is the reflectivity of the plasma surface. The same relation can be obtained by spatially integrating the PMF. The expanding plasma surface is pushed into the target when the light pressure exceeds the plasma pressure ($P> nkT_e$), which is usually the case for high intensity lasers.

%
%
%

The proper description of the PMF should include electron and ion motion at the particular intensity. As the electrons respond instantaneously and ions take time to respond due to their heavy mass, the dynamics are seen on two different timescales. The electrons can be assumed to oscillate with the laser field in a constant ion background and having a slow drift corresponding to the ion motion over the full laser duration. The electrons are pulled out and pushed back in every cycle of the laser pulse, forming relativistic electron jets and generating of higher harmonics from the plasma. Although the ion motion in one optical cycle of the laser pulse is negligible, the ions do move significantly over the whole femtosecond laser pulse, which can be seen from the simulations \cite{wilks1992absorption, macchi2005laser}.

Firstly, for Setup-A a 5mm thick BK7 target was irradiated at different intensities at 45\degree angle of incidence. 
As seen in Fig.~\ref{fig:Setup-A_wavefronts}a, the wavefront becomes increasingly curved, indicating that the plasma surface is pushed towards the target as the intensity increases \textcolor{black}{(note: for better visualization, the plot has been flipped upside down along the Z-axis)}. The lineouts of the plots are shown in Fig.~\ref{fig:Setup-A_wavefronts}c. Due to the PMF, density steepening occurs, which results in the curvature of the plasma surface. After the pulse reflects, the hot plasma begins to expand toward the vacuum (typically assumed to  do so exponentially), establishing a density gradient \cite{rakeeb2025capturing}. 

\begin{figure*}[htbp!]
        \centering
        \includegraphics[width=0.8\linewidth]{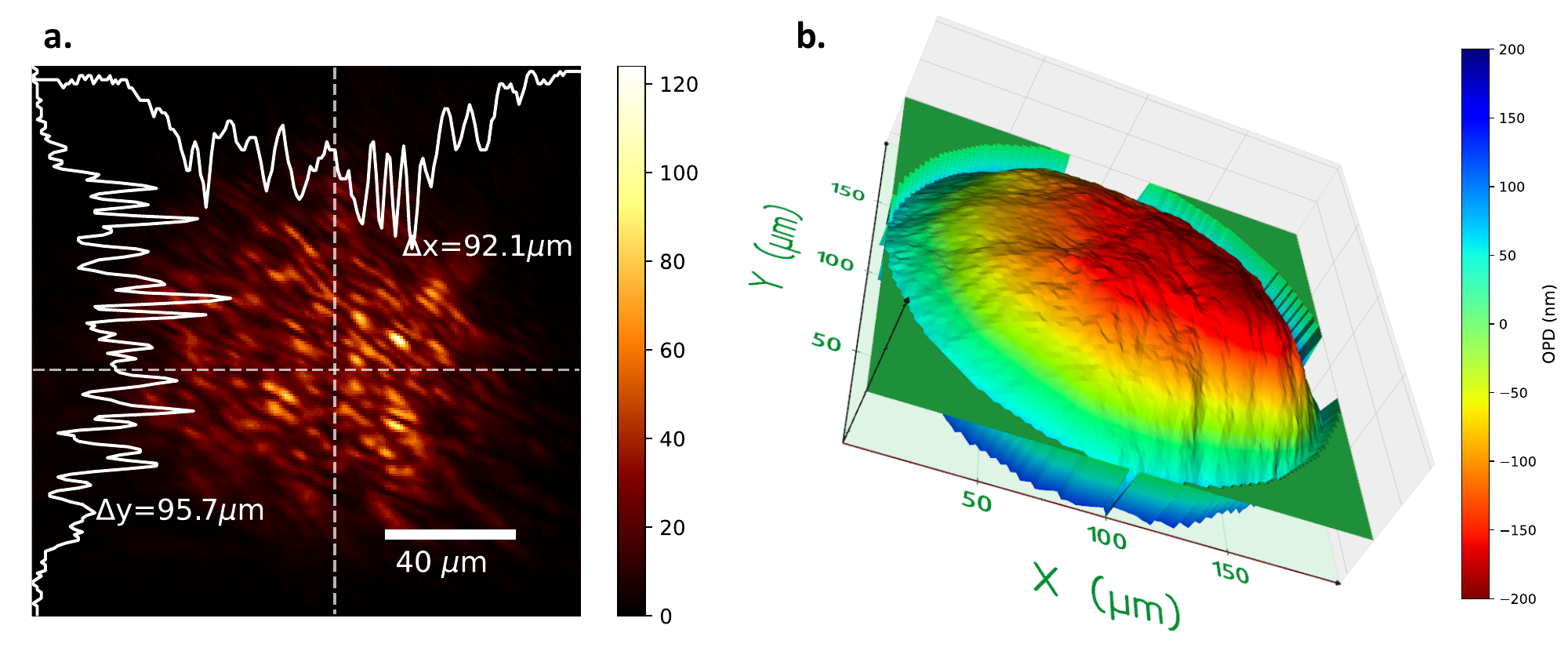}
        
        \caption{\textbf{a.} Experimentally measured focal spot from the beam homogenizer, \textbf{b.} The measured wavefront at highest reachable intensity $4\times 10^{17}$ W/cm$^2$.}
        \label{fig:homogenizer_wf}
\end{figure*} 

\begin{figure*}[htbp!]
        \centering
        \includegraphics[width=1\linewidth]{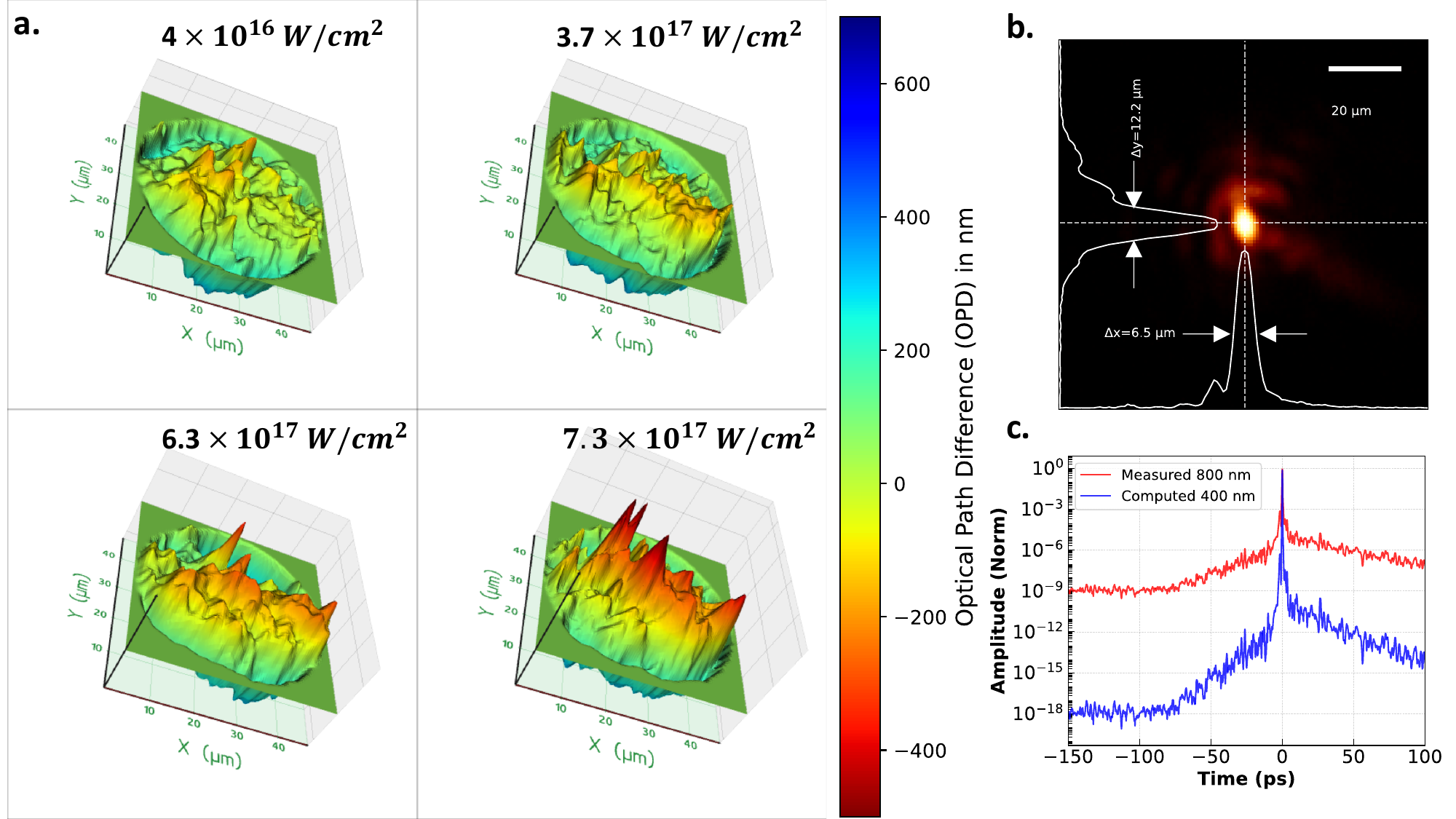}
        
        \caption{\textbf{a)} Measured wavefronts of the extreme contrast pulse. The headings indicate the peak intensities corresponding to each shot, in units of $W/\mathrm{cm}^2$. \textbf{b)} Measured focal spot of the 400~nm pulse at the target surface, showing elongation along the $y$-direction. \textbf{c)} Computed temporal contrast of the second harmonic (400~nm) pulse, derived from the measured contrast of the 800~nm fundamental pulse.}

        \label{fig:400nm_wf_data}
\end{figure*}

\begin{figure}[htbp!]
        \centering
        \includegraphics[width=1\linewidth]{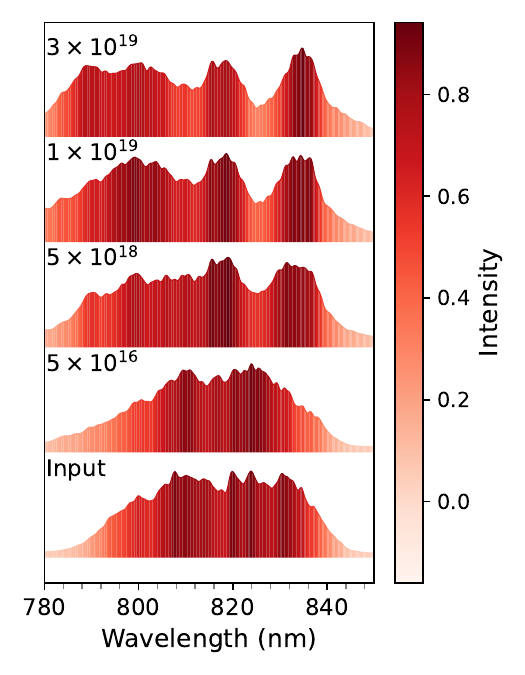}
        
        \caption{Spectrum of the reflected 800~nm pulse. The corresponding intensities are indicated in units of $W/\mathrm{cm}^2$. At higher intensities, the spectrum becomes broadened, and partial absorption is observed near 825~nm.}
        \label{fig:Updated_800nm_spectrum}
\end{figure}

\begin{figure}[htbp!]
        \centering
        \includegraphics[width=1\linewidth]{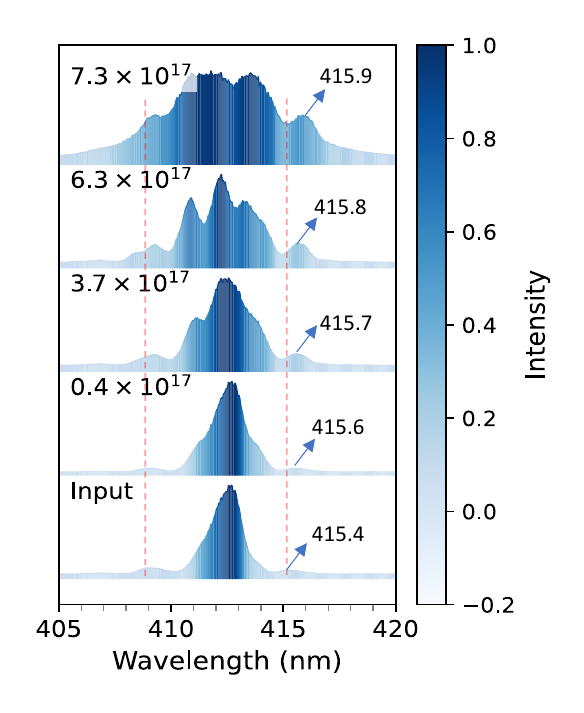}
        
        \caption{Spectrum of the reflected ultrahigh-contrast 400~nm pulse. The corresponding intensities are indicated in units of $W/\mathrm{cm}^2$. The spectral bandwidth increases monotonically with intensity. At the highest intensity, the spectrum broadens to nearly three times its initial width, indicating the potential for generating a more compressible pulse.}
        \label{fig:Updated_400nm_spectrum}
\end{figure}

The experiment was conducted using a high-contrast laser pulse, assuming an exponential density distribution, \(n(x) = n_0 \exp(-x/L)\), where \(n_0\) represents the solid density, and \(L\) is the scale length. The scale length, defined as the position where the density decreases to \(1/e\) of the initial value, significantly influences laser absorption, the energy distribution of heated electrons, and high harmonic generation. The scale length \(L\) was determined\cite{dulat2022subpicosecond} using the frequency domain interferometry technique~\cite{geindre1994frequency}. Fig.~\ref{fig:Setup-A_wavefronts}a presents the measured wavefronts and optical path difference (OPD) maps for various laser intensities, while Fig.~\ref{fig:Setup-A_wavefronts}c shows the corresponding line cuts of the OPD maps.

We use the model developed in \cite{vincenti2014optical} to describe our experimental data. The total push by the laser can be described by electron excursion $x_e(t)$ from the ion surface, $x_i(t)$ which also move significantly over the laser pulse. The total surface deformation is then given by $x_T(t) = x_i(t) + x_e(t) $. The expressions for the $x_i(t)$ and $x_T(t)$ are given by, 

\begin{equation}
x_{i}(t)=2 L \ln \left(1+\frac{\Pi_{0}}{2 L \cos \theta} \int_{0}^{t} a_{L}\left(t'\right) d t'\right)
\end{equation}

\begin{equation}
\resizebox{\columnwidth}{!}{
  $x_{T}(t)=x_{i}(t)+L \ln \left[1+\frac{2 a_{L}(t)(1+\sin \theta)}{2 \pi L / \lambda_{L}} \frac{n_{c}}{n_{0}} e^{-x_{i}(t) / L}\right]$ }
\end{equation}

where $ \Pi_{0}=(R \cos \theta Z m_{e} /2 {A m_{p}})^{1/2}$ with $A$ and $Z$ are ion mass number and average charge state, $m_p$ and $m_e$ are proton mass and electron mass, $\theta$ is the incidence angle, $\lambda_{L}$ is the laser wavelength, and $a_{L}$ corresponds to the laser envelope and is dependent on time and space for a focused laser beam. We fit this model to the experimental data and simulation results, which shows good agreement Fig.~\ref{fig:Setup-A_wavefronts}c. 

\begin{figure}[htbp!]
        \centering
        \includegraphics[width=0.8\linewidth]{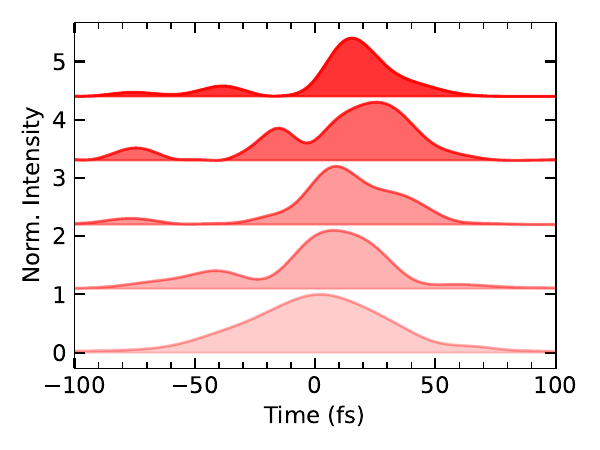}
        
        \caption{Temporal profiles of the reflected fundamental pulse. Intensities are same as in Fig.~\ref{fig:Updated_800nm_spectrum}a in increasing order from bottom to top.}
        \label{FROG_800nm}
\end{figure} 

In recent times surfaces with sub-wavelength structures have attracted great attention as they couple the laser much more efficiently than a planar surface \cite{kahaly2008near,fedeli2016electron}.
To study how sub-wavelength structuring modifies plasma formation and the resulting ponderomotive dynamics, we performed wavefront measurements on a diffraction grating target [Setpu-B]. The grating used in our experiment was a Jobin-Yvon reflective grating with a groove density of 1800 lines/mm, corresponding to a 555\,nm grating period, a 158\,nm groove depth, and a 17.5$^\circ$ blaze angle \cite{kahaly2008near}. Wavefront measurements were performed at multiple laser intensities at angle, 23.5$^\circ$, corresponding to the surface plasmon resonance (SPR) angle of the grating for an 800\,nm laser. At this resonance angle, the incident light couples efficiently into a surface plasmon polariton (SPP) mode, leading to strongly enhanced absorption \cite{kahaly2008near, aparajit2025role}. We know this excites a hotter, denser plasma and therefore it is interesting to see how the critical surface appears for such target. Fig.~\ref{fig:grating_wf_235} presents the reconstructed wavefronts for different intensities at 23.5$^\circ$. A qualitative comparison with the plane target reveals that the grating structure undergoes significantly larger plasma surface deformation near the SPR angle.
The enhanced deformation occurs when the incident field excites a surface plasmon, increasing the local electric field amplitude at the grating ridges and grooves.
This enhanced field leads to a much stronger ponderomotive force, which pushes electrons more effectively away from the high-intensity regions. Consequently, the reflected pulse acquires a strongly curved wavefront that directly maps the underlying plasma surface profile. These observations confirm that structured targets experience stronger ponderomotive reshaping than flat surfaces under similar conditions. 

To further verify the dependence of the ponderomotive force on the laser transverse beam profile, we performed a set of experiments using a flat-top intensity distribution at the target plane [Setup-C]. A beam homogenizer (Holo OR) was introduced in the beamline to reshape the initially Gaussian beam into a near uniform (flat-top) profile, as illustrated in Fig.~\ref{fig:homogenizer_wf}a. 

Due to the homogenization, the resulting focal spot size at the target was larger, approximately $90\,\mu\mathrm{m}$. As a consequence, even the laser at full available energy gave a peak intensity on the target nearly two orders of magnitude lower than that in the standard Gaussian focusing configuration. While this reduction limits the achievable ponderomotive pressure, it allows a controlled investigation of how the spatial beam uniformity influences plasma surface deformation. The wavefront measured at the highest reachable flat-top intensity is presented in Fig.~\ref{fig:homogenizer_wf}b. Compared to the Gaussian case, the plasma surface shows noticeably weaker and more spatially uniform deformation, consistent with the reduced transverse intensity gradients and therefore a lesser ponderomotive driving force.

Additional experiments were performed using an extreme-contrast second harmonic pulse (400~nm) [Setup-D] generated from an LBO crystal \cite{aparajit2021efficient}. The measured pulse profile and its variation with intensity are detailed in Aparajit \textit{et al.} \cite{aparajit2023femtosecond}. The computed temporal contrast and the experimentally measured focal spot are shown in Fig.~\ref{fig:400nm_wf_data}c and Fig.~\ref{fig:400nm_wf_data}b, respectively. The measured wavefronts at different intensities are plotted in Fig.~\ref{fig:400nm_wf_data}a. The ponderomotive dent is observed to deepen with increasing intensity. The elliptical shape of the wavefront closely reflects the focal spot characteristics of the beam (Fig.~\ref{fig:400nm_wf_data}b). 

\begin{figure*}[htp!]
    \centering
    \includegraphics[width=1\linewidth]{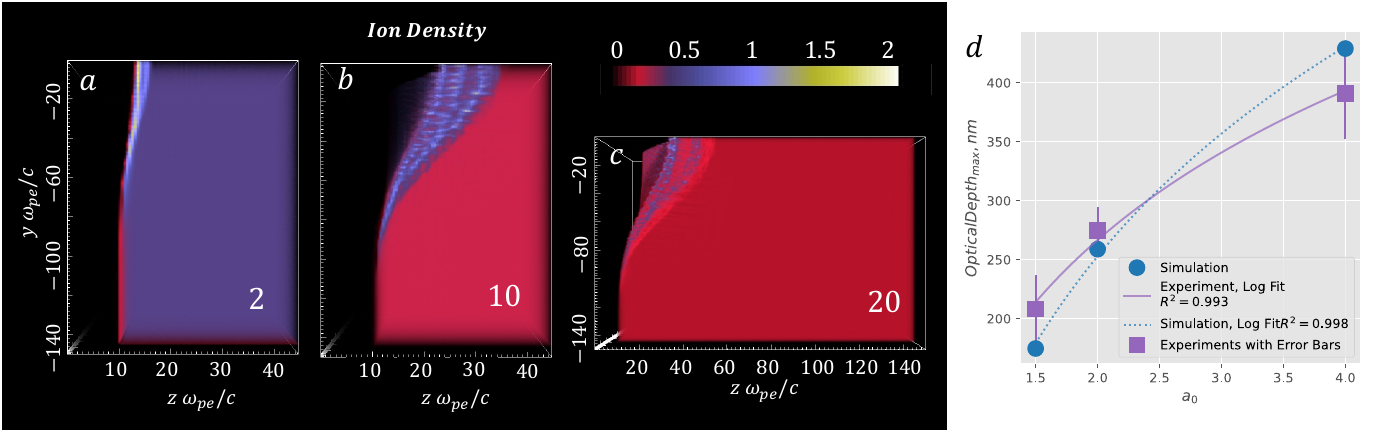}
    \caption{The ion density of the overdense target for non-dimensional laser amplitudes of (a) 2, (b) 10, and (c) 20 is presented after interaction with a relativistic laser. The ponderomotive push results in a cavity whose mean depth increases with amplitude. Additional features such as longitudinal and lateral striations are visible. A comparison of the maximum cavity depth, as measured in the experiment and calculated in the simulation, is shown in (d). The simulation(blue circles) and experimental(purple squares) values show agreement, confirming the logarithmic dependence(simulation fit: dotted blue line).}
    \label{fig:Simulation_ponderomotive_push}
\end{figure*}

\begin{figure}[htp!]
    \centering
    \includegraphics[width=0.8\linewidth]{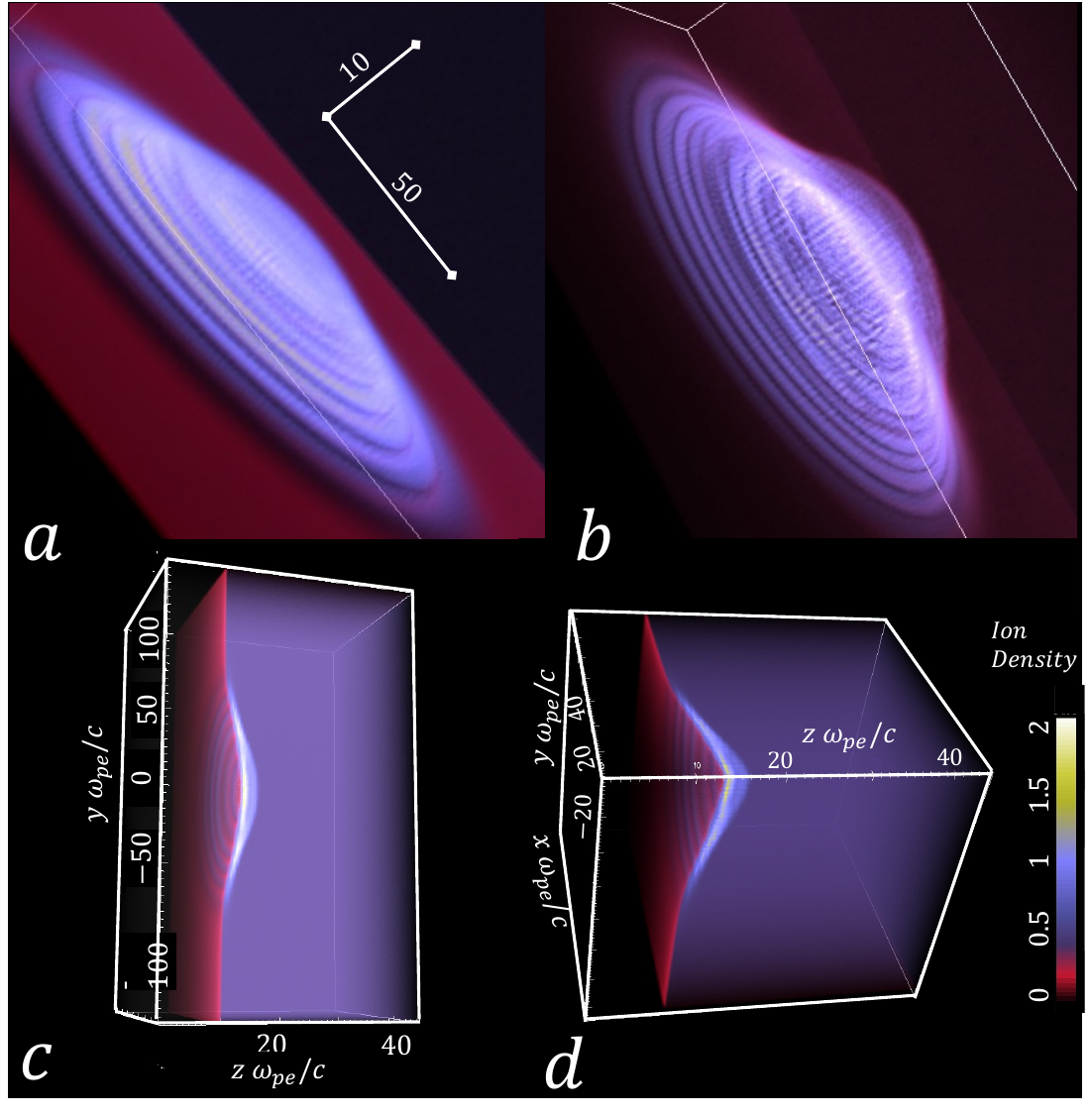}
    \caption{Zoomed view of (a)$a_0 =2$ and  (b) $a_0 =4$ and (c,d) cut section views for $a_0=2$ of the cavity created by the relativistic pulse are presented. }
    \label{fig:ion-dens}
\end{figure}

 A part of the reflected light is also sent to a spectrometer to analyze the spectral behavior of the plasma mirror. For Setup-A the reflected spectrum is shown in Fig.~\ref{fig:Updated_800nm_spectrum}. The interaction with the plasma surface leads to heavy modification of the incident laser spectrum. At high intensities, significant spectral broadening is observed, accompanied by strong absorption around 825~nm. As the incident intensity is reduced, the reflected spectrum gradually approaches the original laser spectrum. The observed spectral broadening can be attributed to self-phase modulation arising from the rapid temporal variation of the plasma refractive index, suggesting the possibility of pulse compression with appropriate chirp compensation. In this case, substantial pulse-shape deformation is also observed, as presented in Fig.~\ref{FROG_800nm}.
 
The spectrum of the reflected extreme-contrast 400~nm pulse in Setup-D is shown in Fig.~\ref{fig:Updated_400nm_spectrum}. A clear intensity-dependent spectral modulation and broadening is evident. At the highest intensity, the spectral width increases to nearly three times its initial value, indicating the potential for generating a significantly compressed pulse. The redshift of the spectral peak at 415.4~nm further supports the presence of an increasing plasma expansion velocity into the solid target. Similar behavior is not clearly observed in Setup-A, primarily because the fundamental 800~nm pulse already possesses a significantly broad bandwidth.

To measure the temporal profile of the reflected laser pulse after interaction in setup-A, the collected light was collimated and directed to a GRENOUILLE (Frequency-Resolved Optical Gating) device. Due to the collection optics and the chamber window, the pulse undergoes temporal stretching before reaching the detector; nevertheless, the measurement provides valuable qualitative insight into the pulse evolution at different intensities. Fig.~\ref{FROG_800nm} shows the retrieved temporal profiles of the reflected pulse. The interaction with the plasma mirror leads to severe pulse distortion, with the appearance of multiple temporal sub-structures at increasing intensities, while the duration of the central peak progressively decreases.
\section{Simulation}

To complement our experimental observations, we performed 3D3V and 2D3V Particle-In-Cell (PIC) simulations using the massively parallel, fully relativistic, open-source OSIRIS 4.0 framework \cite{fonseca2002osiris}. These simulations provide a detailed understanding of plasma dynamics at resolutions and time scales inaccessible in experiments, offering insights into energy deposition, charge transport, and harmonic generation.

The simulations were carried out in a domain of size $300 \times 300 \times 150$ cells, with a grid resolution of $dx =dy=dz=0.5 \omega_{pe} /c $. A linearly-polarized (along $\hat{y}$) laser pulse of wavelength $( \lambda = 800 ) nm$, duration $10 fs$, and spot size $1 \mu m$ was incident normally (\( 0^\circ \) angle of incidence) on a fully ionized plasma. The laser intensity was varied with normalized field strengths $( a_0 = 0.5, 1.0, 1.5,2,10,20 )$ to study the effects of intensity on plasma evolution ($a_0 = 0.86\lambda_{L}[\mu m]\sqrt{I_0 [10^{18} W/cm^{2}]}$. for $\lambda_L = 0.8 \mu m, I_0 [10^{18} W/cm^2] = 2.14, a_0 = 1.01$  ).

The plasma target was modeled at density  \( 50n_c \), consisting of electrons and static neutralizing ions. The ion-to-electron mass ratio was chosen as \( m_i/m_e = 25 \). The simulation parameters, including laser and plasma properties, are summarized in  Table \ref{table1} .

\begin{table}[h]
    \centering
    \caption{\label{table2} Simulation Parameters}
    \resizebox{\columnwidth}{!}{
    \begin{tabular}{l l l l}
    \hline
    \hline
    Parameters                &                   &Standard Units                               &Normalized Units\\ 
    \hline
    \textit{Laser}  \\
    Wavelength                & \( \lambda_{L} \) & 800 nm                                      &\( 44 [c/\omega_{pe}] \)\\
    Frequency                 & \( \omega_{L} \)  & \( 2.35 \times 10^{15} \) rad/s             & \( 0.14 [\omega_{pe}] \)\\
    Intensity                 & \( I_{o} \)      & \( 5.37 \times10^{17} \) W/cm\(^2\)        & \( a_{o} = 0.5 \)\\
    \hline
    \textit{Plasma}\\
    Electron Plasma frequency & \( \omega_{pe} \) & \( 1.66 \times 10^{16} \) rad/s             &\( 1 [\omega_{pe}] \)\\
    Critical density          & \( n_{c} (\omega_L) \) & \( 1.73 \times 10^{27} \) 1/m\(^3\)   &\\
    Electron number density   & \( n_{e} \)       & \( 50 n_{c}, 8.67 \times 10^{28} \) 1/m\(^3\) &\\
    Ion number density        & \( n_{i} \)       & \( 8.67 \times 10^{28} \) 1/m\(^3\)         &\\
    Mass of electrons         & \( m_e \)         & \( 9.109 \times 10^{-31} \) kg             &\\
    Mass of ions              & \( m_i \)         & \( 25  m_{e} \) \\
    \hline
    \hline
    \end{tabular}
    }
\end{table}

The {depth of the plasma surface depression} due to the ponderomotive push exerted by the laser field was analyzed as a function of \( a_0 \), as shown in  Fig.~\ref{fig:Simulation_ponderomotive_push} . The results indicate a monotonic increase in penetration depth with increasing laser intensity, consistent with theoretical predictions of ponderomotive scaling.
The modification of plasma density due to laser interaction is illustrated in  Fig.~\ref{fig:ion-dens} , showing snapshots of electron, ion, and charge densities. The charge separation induced by the laser field results in the formation of a characteristic plasma cavity at higher intensities.
To analyze the impact of laser intensity and plasma density on the reflected electromagnetic field, we performed a Fourier analysis of the transverse field component. Notably, the reflected field exhibits odd harmonics, a signature of relativistic plasma boundary oscillations, which arise due to nonlinear plasma interactions\cite{Baeva2006,Ganeev2007}.


Our {PIC simulations validate the experimental findings while providing deeper insights into laser-plasma interactions. The observed trends in ponderomotive push, density modification, and spectral modulation} highlight the importance of laser intensity and plasma density in determining the interaction dynamics. These results serve as a benchmark for future high-intensity laser experiments and theoretical investigations.

\section{Conclusion}

We have reported direct measurement of the spatial, spectral, and temporal characteristics of light reflected from relativistic PMs, in agreement with both particle-in-cell (PIC) simulations and theoretical models. We present a direct experimental measurement of the shape of the critical electron density surface in different kind of high intensity laser excited  plasma on a solid surface for two different laser wavelengths, providing clear evidence of the ponderomotive push.
The average velocity of the deformation observed for Setup-A was $0.043c$ at the intensity of $3 \times 10^{19}~W/\mathrm{cm}^2$, in line with the theoretical value of $0.025c$ at $1.5 \times 10^{18}~W/\mathrm{cm}^2$ reported by Wilks \textit{et al.} (1992) \cite{wilks1992absorption}.

The behavior of PMs was investigated using high- and extreme-contrast pulses. Significant spectral broadening and pulse shape modification are observed, arising from the deformation of the PM and various nonlinear plasma dynamics. As the conventional metal-coated or dielectric mirrors are unsuitable at relativistic intensities due to their low damage thresholds, relativistic PMs provide a promising alternative for reaching extreme intensities and exploring quantum electrodynamics (QED) regimes. The curvature induced to the reflected beam by the PM can be utilized in the far field for multiple applications, such as controlling the divergence of attosecond pulses generated during the interaction or focusing the reflected beam.

\section{Acknowledgements}

AD acknowledges support from the Anusandhan National Research Foundation (ANRF) of the Government of India through core grant CRG/2022/002782 as well as a J C Bose Fellowship grant JCB/2017/000055. 
GRK acknowledges major support for this research from the grant “Physics and Astronomy (Project Identification No. RTI4002) Department of Atomic Energy, Tata Institute of Fundamental Research” and partially from the grant JBR/2021/00039 of the Anusandhan National Research Foundation (ANRF), both of the Government of India.


\bibliography{reference.bib}
\end{document}